\documentclass{elsart}

\usepackage{epsfig}

\usepackage{amssymb}
\usepackage{amsmath}

\begin{document}

\begin{frontmatter}



\title{The fully entangled fraction as an inclusive measure of
entanglement applications}

\author{J. Grondalski\corauthref{cor1}},
\ead{jcat@t4.lanl.gov}
\corauth[cor1]{Corresponding author.}
\author{D.M. Etlinger\thanksref{now}},
\thanks[now]{Present address:  University of Rochester, 
Rochester, NY 14627.}
\author{D.F.V. James}

\address{Theory Division, T-4, Los Alamos National Laboratory, \\
Los Alamos, New Mexico 87545, USA}

\begin{abstract}

Characterizing entanglement in all but the simplest case of a two qubit
pure state is a hard problem, even understanding the relevant experimental
quantities that are related to entanglement is difficult.  It may not be
necessary, however, to quantify the entanglement of a state in order to
quantify the quantum information processing significance of a state.  It
is known that the fully entangled fraction has a direct relationship to
the fidelity of teleportation maximized under the actions of local unitary
operations.  In the case of two qubits we point out that the fully
entangled fraction can also be related to the fidelities, maximized under
the actions of local unitary operations, of other important quantum
information tasks such as dense coding, entanglement swapping and quantum
cryptography in such a way as to provide an inclusive measure of these
entanglement applications.  For two qubit systems the fully entangled
fraction has a simple known closed-form expression and we establish lower
and upper bounds of this quantity with the concurrence.  This approach is
readily extendable to more complicated systems.

LA-UR-02-1487

\end{abstract}

\end{frontmatter}

\section{Introduction}

A pure quantum state is entangled if it is impossible to factorize into a
tensor product of states for the separate systems (e.g., the singlet state
of two spin-$\frac{1}{2}$ particles, $(1/\sqrt{2})(|01 \rangle - |10
\rangle)$, is entangled).  This property, originally introduced to sharpen
discussions of foundational issues in quantum theory \cite{1}, has been
studied extensively with regard to nonlocal quantum correlations \cite{2}
indicated by the observed violation of Bell's inequality \cite{3}.  In the
past decade, the focus of entanglement studies has shifted toward
applications which use the nonclassical features of quantum systems to
surpass classical limitations on communications and computation.  Such
applications are part of the emerging field of {\em quantum information}
\cite{4} and include quantum cryptography \cite{5}, dense coding \cite{6},
teleportation \cite{7}, entanglement swapping \cite{8}, and quantum
computation \cite{9}.

Due to this recent interest in quantum entanglement applications, the
characterization of entanglement in a mixed bipartite system has become an
intensely studied problem.  In general, mixed states are entangled if it
is impossible to represent the density operator as an incoherent sum of
factorizable pure states \cite{10}.  There are a number of measures of
entanglement for a bipartite system.  Three closely related measures are
the entanglement of formation, the entanglement of distillation, and the
concurrence.  The entanglement of formation is defined as the least number
of maximally entangled states required to asymptotically prepare a mixed
state $\rho$ with local operations and classical communications \cite{11}
and the entanglement of distillation is defined as the asymptotic yield of
maximally entangled states that can be extracted from $\rho$ with local
operations and classical communications \cite{11}.  The concurrence
\cite{12} is monotonically related to the entanglement of formation, and
therefore an equally valid measure of entanglement, but is the only
measure described here that provides a closed expression for the simplest
case of a two qubit bipartite system \cite{13}.  Relative entropy
\cite{14} measures entanglement by considering the ability to distinguish
$\rho$ from all separable states and negativity \cite{15} quantifies the
degree to which the eigenvalues of the partial transpose fail to satisfy
the partial transpose separability condition \cite{16}.  To be sure all of
these entanglement measures can be computed, like any physical quantity in
quantum mechanics, from knowledge of the density matrix which can be found
experimentally with tomography \cite{17}, but their relation to
experimental consequences are indirect at best.  For example, a two qubit
mixed state described by an ensemble of partially entangled states can
always be distilled, in a non-unique fashion, into a smaller ensemble of
maximally entangled states which can in turn be used for useful quantum
information processing \cite{18}.

\begin{figure}[t]
\begin{center}
\epsfig{file=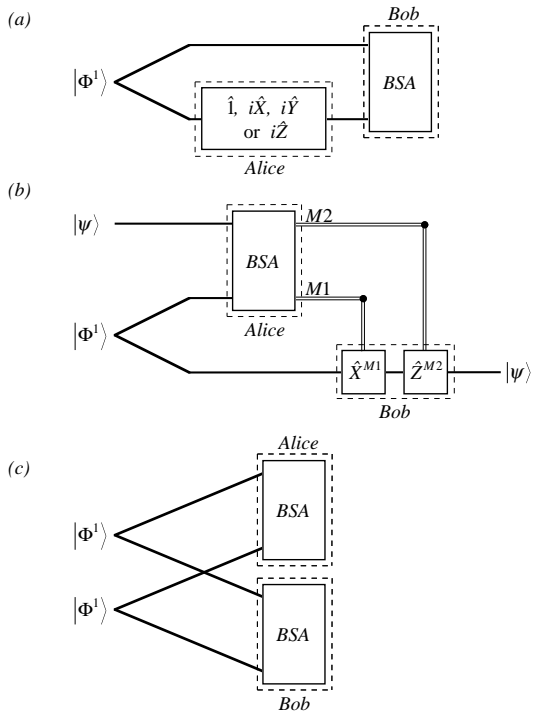,width=7cm}
\caption{Circuit diagram representation for: (a) Dense coding:  Alice
apples one of four unitaries $\{\hat{1},i\hat{X},i\hat{Y},i\hat{Z}\}$ to
her qubit which Bob can read out with a Bell state analysis (BSA).  (b)
Teleportation:  Alice teleports an unknown quantum state $|\psi\rangle$ by
sending the result of a Bell state analysis $\{M1,M2\}$ to Bob who
transforms his qubit into $|\psi\rangle$ conditioned on this information.  
(c) Entanglement swapping:  Alice projects Bob's two particles into a
maximally entangled state via a Bell state analysis on her two qubits.  
See Ref.~\cite{4} for a complete description of quantum circuit diagrams.}
\end{center}
\end{figure} 

Modern conventional wisdom holds that characterizing entanglement in all
but the simplest of cases is a hard problem.  Even understanding the
relevant experimental quantities that are related to entanglement is
difficult.  It may not be necessary, however, to quantify the entanglement
of a state in order to quantify the quantum information processing
significance of a state.  For example, Horodecki et.\ al.\ \cite{19}
demonstrated that the maximum teleportation fidelity for a general two
qubit system is given by,
\begin{equation}
F_{T}^{max} = \frac{1}{3} \: \bigl( 1 + 2
\: F \bigr).  
\end{equation}
where $F$ is the fully entangled fraction \cite{11} and is defined as the 
overlap between a mixed state $\hat{\rho}$ and a maximally
entangled state $|\Phi\rangle$ maximized over all $|\Phi\rangle$, 
\begin{equation}
F = \underset{|\Phi\rangle}{\mbox{max}}\{\langle\Phi|
\hat{\rho}|\Phi\rangle\}.
\end{equation}
Unlike entanglement, the fully entangled fraction does have a clear
experimental interpretation as the optimal ability of a state to teleport
and it is clear that the degree to which fully entangled fraction is
greater than $1/2$ ($F_{T} > 2/3$) can be used to quantify the teleporting
ability of a state over the best ``classical teleportation'' protocols.  
This suggests that it may be possible to define a {\it measure of
entanglement applications} directly.  Such a mathematical quantity may be
just as useful as a true entanglement measure, but more practical from a
theoretical standpoint.  It is natural to wonder whether the fully
entangled fraction is such a quantity, that is, can it be used to measure
the general quantum information significance of a state.  To answer this
question for the case of two qubits, in Sec.~2, we examine the
relationship between the fully entangled fraction and the fidelities of
all two qubit applications which have been experimentally demonstrated to
date:  dense coding, teleportation, entanglement swapping, and quantum
cryptography (Bell inequalities).  We consider these applications with a
general two qubit mixed state in place of the standard maximally entangled
pure state and find that the fully entangled fraction does indeed quantify
the quantum processing significance of dense coding, teleportation,
entanglement swapping, and quantum cryptography (violating Bell's
inequality) in an inclusive sense.  We are taking ``inclusive'' to mean
that a non-zero value indicates that a state can perform at least one of
these applications better than allowed ``classically'' and a zero value
indicates that a state cannot perform any of these applications better
than allowed ``classically''.  Similar approaches have examined continuous
variable teleportation \cite{20} and Bell inequality experiments
\cite{21}.  The fully entangled fraction has a simple closed-form analytic
expression, in the case of two qubits, which we rederive in Sec.~3 under
the present context.  In Sec.~4 we establish upper and lower bounds
between the fully entangled fraction and the only measure of entanglement
described above with a closed-form expression for a general two-qubit
state, the concurrence, and in Sec.~5 we conclude and discuss
generalizations of these ideas to more complicated systems.

\section{The relation between two-qubit applications and the maximally
entangled fraction}

\subsection{Dense Coding}

The relationship between the fully entangled fraction and dense coding
\cite{6} (See Fig.~1a) is clearest.  In this entanglement application
Alice and Bob each receive one qubit of a maximally entangled state,
$|\Phi^{1}\rangle \equiv \sqrt{1/2}(|00\rangle + |11\rangle)$, where the
first entry denotes Bob's qubit and the second denotes Alice's
qubit.  Alice can encode 2 bits of information in four orthogonal
states by applying one of four local unitaries {\em solely} to her own
qubit,
\begin{eqnarray}
\hat{1} \otimes \hat{1}|\Phi^{1}\rangle & = & 
|\Phi^{1}\rangle \nonumber \\
\hat{1} \otimes i\hat{X} |\Phi^{1}\rangle & = &
|\Phi^{2}\rangle \equiv i \:
\frac{|01\rangle + |10\rangle}{\sqrt{2}} \nonumber \\
\hat{1} \otimes i\:\hat{Y} |\Phi^{1}\rangle & = &
|\Phi^{3}\rangle \equiv 
-\frac{|01\rangle - |10\rangle}{\sqrt{2}} \nonumber \\
\hat{1} \otimes i\hat{Z} |\Phi^{1}\rangle & = &
|\Phi^{4}\rangle \equiv i \: 
\frac{|00\rangle - |11\rangle}{\sqrt{2}},
\end{eqnarray}
where $\{\hat{X},\hat{Y},\hat{Z} \}$ are the Pauli operators \cite{4}
and $|\Phi^{j}\rangle$ $(j = 1,2,3,4)$ are the so-called ``magic basis''
states \cite{11,12}, an orthonormal set of maximally entangled states
with a convenient phase convention.  After receiving one
qubit from Alice, Bob can read out two
bits of information with a fidelity of one by performing a Bell state
analysis on his two particles.  To measure entanglement with this protocol
the maximally entangled state $|\Phi^{1}\rangle$ is replaced with an
arbitrary two qubit state $\hat{\rho}$ and the dense coding fidelity is
defined as an average over the four possible outcomes,
\begin{eqnarray}
F_{DC} & = & \frac{1}{4} \Bigl(
\langle\Phi^{1}|\hat{\rho}|\Phi^{1}\rangle + 
\langle\Phi^{2}|(\hat{1} \otimes i\hat{X}) \hat{\rho}
(\hat{1} \otimes i\hat{X})^{\dagger}|\Phi^{2}\rangle \nonumber \\
& &+\langle\Phi^{3}|(\hat{1} \otimes i\hat{Y}) \hat{\rho}
(\hat{1} \otimes i\hat{Y})^{\dagger}|\Phi^{3}\rangle \nonumber \\
& &+ \langle\Phi^{4}|(\hat{1} \otimes i\hat{Z}) \hat{\rho}
(\hat{1} \otimes i\hat{Z})^{\dagger}|\Phi^{4}\rangle \Bigr).
\end{eqnarray}
Using the definition of the $|\Phi^{j}\rangle$ basis states, the dense
coding fidelity reduces to the fidelity of $\hat{\rho}$ relative to a
single maximally entangled state,
\begin{equation}
F_{DC} = \langle\Phi^{1}|\hat{\rho}|\Phi^{1}\rangle.
\end{equation}
However, it is natural to expect that, experimentally, one should attempt
to maximize the utility of the state by choosing the best possible local
coordinate basis in which to carry out the experiment.  Thus, the
intrinsic capabilities of the state should take this into account.  
Mathematically, this is expressed by maximizing the quantity
$\langle\Phi^{1}|\hat{\rho}|\Phi^{1}\rangle$ over all possible local
unitary operators, viz., 
\begin{equation} 
F_{DC}^{max}  = \underset{\hat{U}_{A},\hat{U}_{B}}{\mbox{max}}
\{\langle\Phi^{1}| (\hat{U}_{A}\otimes \hat{U}_{B})^{\dagger} \hat{\rho}
(\hat{U}_{A}\otimes \hat{U}_{B}) |\Phi^{1}\rangle \}.
\end{equation}
Because all maximally entangled states are related under local unitary
operations, this is equivalent to maximizing
$\langle\Phi|\hat{\rho}|\Phi\rangle$ over all maximally entangled states
$|\Phi\rangle$ which is just the fully entangled fraction,
\begin{equation}
F_{DC}^{max} = F.
\end{equation}
It is clear that the maximum fidelity for dense coding 
($F_{DC}^{max}=1$) occurs when $\hat{\rho}$ is maximally entangled and
the maximum fidelity for a separable state ($F_{DC}^{max}=1/2$) occurs
when $\hat{\rho}$ is pure.

\subsection{Teleportation}

For pedagogical reasons we next examine teleportation \cite{7} (See
Fig.~1b), whose relationship to the fully entangled fraction  was
first worked out by Horodecki et.\ al.\ \cite{19} The goal of
teleportation is to use
a maximally entangled pair of qubits to transmit an arbitrary quantum
state from one point to another
with the communication of only two classical bits.  Briefly, Alice has a
qubit (particle 1) in an unknown quantum state $|\psi\rangle_{1} =
\cos(\theta/2)\,|0\rangle + \sin(\theta/2)e^{i \phi}\,|1\rangle $ 
($0< \theta < \pi, 0< \phi < 2\pi$) and
Alice and Bob again share a maximally entangled state
$|\Phi^{1}\rangle_{23}$ (particles 2 and 3 respectively).  
Alice performs a Bell state analysis on her two qubits (particles 1 and 2)
measuring one of four possible outcomes
$\{M1,M2\} \in \{0,1\}$.  Using two bits of classical information she
informs Bob of the outcome and he applies the unitary transformation
$\hat{Z}^{M2}\hat{X}^{M1}$, transforming
his qubit into Alice's original quantum state with a fidelity of
unity.  Suppose that, instead of the maximally entangled state
$|\Phi^{1}\rangle$, we attempt teleportation using an arbitrary two
qubit state $\hat{\rho}$.  Following Popescu \cite{22}, we
define the teleportation fidelity as an ensemble average over all input
states $|\psi\rangle$, 
\begin{equation}
F_{T} = 
\frac{1}{4 \pi} \int_{0}^{2 \pi} \int_{0}^{\pi} 
f(\theta,\phi) \: \sin(\theta) d\theta d\phi,
\end{equation}
where $f(\theta,\phi) = \langle \psi| \hat{\rho}_{out} |\psi
\rangle$.  This quantity can be considered a measure of the usefulness of
$\hat{\rho}$ for performing teleportation.  It is easiest to compute by
exchanging the measurements and control operations \cite{23} so that
the Bell basis transformation is followed by a controlled-NOT between
particles 2 and 3 and a controlled-Z between particles 1 and 3 (See
Ref.~\cite{4}, Ch.~4 for a description of these quantum
gates).  The trace over Alice's system can be performed and the integral
above can be computed to give,
\begin{equation}
F_{T}  =  \frac{1}{3} \: \bigl( 1 + 2
\: \langle\Phi^{1}|\hat{\rho}|\Phi^{1}\rangle \bigr).
\end{equation}
Again, maximizing this quantity over all over all possible local
unitary operators gives,
\begin{equation}
F_{T}^{max} = \frac{1}{3} \: \bigl( 1 + 2
\: F \bigr).
\end{equation}

\subsection{Entanglement Swapping}

The relationship between the fully entangled fraction and entanglement
swapping \cite{8} (See Fig.~1c) is similar to the case of dense
coding.  In this entanglement application, there are two pairs of
maximally entangled states in a direct product state 
$|\Phi\rangle_{1234}=|\Phi^{1}\rangle_{12}\otimes|\Phi^{1}\rangle_{34}$
(where $1,2,3,$ and $4$) label the particles respectively.  If Alice
receives particles $1$ and $3$ and Bob receives particles $2$ and $4$, the
state can be reexpressed in this basis as,
\begin{eqnarray}
|\Phi\rangle_{1234} &=& \frac{1}{2}
\Bigl( |\Phi^{1}\rangle_{13} \otimes |\Phi^{1}\rangle_{24} -
|\Phi^{2}\rangle_{13} \otimes |\Phi^{2}\rangle_{24} \nonumber \\ 
& & + |\Phi^{3}\rangle_{13} \otimes |\Phi^{3}\rangle_{24} -
|\Phi^{4}\rangle_{13} \otimes |\Phi^{4}\rangle_{24} \Bigr).
\end{eqnarray}
A Bell measurement by Alice (Bob) will project Bob's (Alice's) particle
into a maximally entangled state despite the fact that the two particles
have never interacted in the past.  Replacing either maximally entangled
state with an arbitrary density matrix $\hat{\rho}$ and making use of
symbolic manipulation software, the fidelity of entanglement swapping can
be defined similarly to dense coding as an average over the fidelities of
the four possible outcomes with a similar result,
\begin{equation}
F_{ES}  = \langle\Phi^{1}|\hat{\rho}|\Phi^{1}\rangle.
\end{equation}
Once again, maximizing this quantity over all over all possible local
unitary operators gives,
\begin{equation}
F_{ES}^{max} = F.
\end{equation}

\subsection{Quantum cryptography (Bell inequalities)}

Last, we examine the relationship between the fully entangled fraction and
Bell inequality experiments, which occur, for example, in the Ekert
protocol for secure key distribution \cite{5}.  The standard Bell
correlation function \cite{24} is given by, \begin{eqnarray} B & = & \Big|
\mbox{Tr} \Big\{ \hat{S}_{1}(\phi_{1}) \hat{S}_{2}(\phi_{2}) \hat{\rho} -
\hat{S}_{1}(\phi_{1}) \hat{S}_{2}(\phi_{2}') \hat{\rho} \nonumber \\ & &
+\hat{S}_{1}(\phi_{1}') \hat{S}_{2}(\phi_{2}) \hat{\rho}
+\hat{S}_{1}(\phi_{1}') \hat{S}_{2}(\phi_{2}') \hat{\rho} \Big\} \Big|,
\end{eqnarray} where $\hat{S}_{j}(\phi_{j}) = \cos(\phi_{j}) \hat{Z}_{j} +
\sin(\phi_{j}) \hat{X}_{j}$.  The Bell inequality is given by $B \leq 2$
and is violated when $B > 2$.  The detectors are set to their optimal
values $\{ \phi_{1}=0,\phi_{1}'=\pi/2,\phi_{2}=\pi/4,\phi_{2}'=3\pi/4 \}$
such that the violation is maximum for the maximally entangled state
$|\Phi^{1}\rangle$.  This state is then replaced with a general state
$\hat{\rho}$ and the Bell correlation function as a function of
$\hat{\rho}$ is given by, \begin{eqnarray} B(\hat{\rho}) & = & \sqrt{2}
\big| \mbox{Tr} \big\{ (\hat{X}\otimes\hat{X} + \hat{Z}\otimes\hat{Z})  
\hat{\rho} \big\} \big| \nonumber \\ & = & \sqrt{2} \big| \sum_{j} \langle
\Phi^{j}| (\hat{X}\otimes\hat{X} + \hat{Z}\otimes\hat{Z})  \hat{\rho}
|\Phi^{j}\rangle \big| \nonumber \\ & = & 2 \sqrt{2} \: |\langle
\Phi^{1}|\hat{\rho}|\Phi^{1}\rangle - \langle
\Phi^{3}|\hat{\rho}|\Phi^{3}\rangle |.  \end{eqnarray} It is clear that
the normalized expression $B/2\sqrt{2}$, maximized over all local
unitaries operating on the separate subsystems, will always be less than
or equal to the fully entangled fraction.  Therefore, a fully entangled
fraction greater than $1/2$ is a sufficiency condition for violating
Bell's inequality.  Munro et.\ al.\ considered a similar situation by
maximizing Bell correlations over all possible detector orientations
$\{\phi_{1},\phi_{1}',\phi_{2},\phi_{2}'\}$ \cite{21}.  We have verified
numerically that this quantity is also always less than the fully
entangled fraction by searching $500,000$ random states weighted toward
higher concurrences (we explain how this is done in Sec.~4).  This result
is not unexpected due to the fact that it is well known that there exist
mixed states which can teleport arbitrary quantum states better than any
classical protocol, yet fail to violate standard Bell inequalities
\cite{22}.

It is interesting to note that in general, measures which maximize the
overlap between a fiducial pure state and an input state with respect to a
local basis, viz.,
\begin{equation}
F(|\psi_{f}\rangle,\hat{\rho}) =
\underset{\hat{U}_{A},\hat{U}_{B}}{\mbox{max}}\{\langle\psi_{f}|
(\hat{U}_{A}\otimes \hat{U}_{B})^{\dagger} \hat{\rho}
(\hat{U}_{A}\otimes \hat{U}_{B}) |\psi_{f}\rangle \},
\end{equation} 
have the largest difference in fidelities between a maximally entangled
state and a separable pure state,
\begin{equation} 
\Delta = F(|\psi_{f}\rangle,|\Phi\rangle\langle\Phi|) - 
F(|\psi_{f}\rangle,|uv\rangle\langle uv|), 
\end{equation}
when $|\psi_{f}\rangle$ is
maximally entangled.  This can be seen by writing the fiducial state in
a Schmidt decomposition $|\psi_{f}\rangle = \bigl( \hat{U}_{A} \otimes
\hat{U}_{B} \bigr) \bigl( \cos(\theta/2)|00\rangle + 
\sin(\theta/2)|11\rangle \bigr)$ \cite{25}, maximizing each term in
Eq.~(17) separately over the local unitary operators, and then maximizing
$\Delta$ with respect to $\theta$ to show that the maximum
occurs for $\theta=\pi/2$.  It is physically intuitive that this statement
will also be true if this measure is generalized to fiducial mixed
states.  Although not a rigorous proof, this suggests that the
fully entangled fraction is the ``best'' quantifier of entanglement
applications in the sense of being the most inclusive.

\section{A simple closed-form expression for the fully entangled fraction}

In Sec.~2 we deduced that the fully entangled fraction can be physically
interpreted as an inclusive measure of entanglement applications.  
That is, when $F$ is greater than $1/2$ a mixed state can at least perform
dense coding, teleportation, or entanglement swapping with a fidelity that
is better than any separable state using classical protocols.  It is clear
that this quantity is invariant under local unitary operators, which can
be viewed passively as a basis transformation, but not under local
non-unitary operators (e.g., projective measurements and dissipation)
\cite{11,26}.  These non-unitary operators can invoke irreversible changes
in a state that are less useful for understanding the intrinsic
properties of a quantum state.  In light of this result we briefly reprise
here the derivation of a closed-form expression, first derived by Bennett
et. al. \cite{11}, of the fully entangled fraction in the case of an
arbitrary state of two qubits.  We take as our starting point the fully
entangled fraction as expressed by Eq.~(6).  This expression can be
simplified by using a property of maximally entangled states,
$\hat{U}_{A}\otimes \hat{U}_{B}|\Phi^{1}\rangle = \hat{1}\otimes
\hat{U}_{B} \hat{Y} \hat{U}_{A}^{\dagger} \hat{Y} |\Phi^{1}\rangle$, and
redefining the optimizing unitary $\hat{U} = \hat{U}_{B} \hat{Y}
\hat{U}_{A}^{\dagger} \hat{Y}$, so that this expression involves only a
single maximization over a local unitary operator,
\begin{equation}
F = \underset{\hat{U}}{\mbox{max}}\{\langle\Phi^{1}|
(\hat{1}\otimes \hat{U})^{\dagger} \hat{\rho}
(\hat{1}\otimes \hat{U}) |\Phi^{1}\rangle \}.
\end{equation}
Expanding $|\Phi\rangle = (\hat{1}\otimes \hat{U}) |\Phi^{1}\rangle$
in a Pauli basis and making use of the basis states defined in Eq.~(1),
\begin{eqnarray}
|\Phi\rangle &=& \hat{1}\otimes\bigl( x_{1} \: \hat{1} +i x_{2}
\: \hat{X} + i x_{3} \: \hat{Y} +
i x_{4} \:  \hat{Z} \bigr) |\Phi^{1}\rangle \nonumber \\
&=& \sum_{n=1}^{4} x_{n} \: |\Phi^{n}\rangle,
\end{eqnarray}
allows one to represent an arbitrary maximally entangled state by four
real parameters $x_{n}$ ($n=1,2,3,4$) that satisfy 
$g(x_{1},x_{2},x_{3},x_{4}) \equiv x_{1}^2 + x_{2}^2 + x_{3}^2 + x_{4}^2 =
1$.  Inserting this expression into Eq.~(18) gives,
\begin{equation}
F(x_{1},x_{2},x_{3},x_{4}) = \sum_{n,m=1}^{4} M_{n,m} \: x_{n} x_{m},
\end{equation}
where $M_{n,m} = \langle \Phi^{n}|\hat{\rho}|\Phi^{m}\rangle$.  The
extrema condition is found by including the constraint with an 
undetermined Lagrange multiplier $\lambda$,
\begin{equation}
\frac{\partial}{\partial x_{k}} \Bigl\{ 
F(x_{1},x_{2},x_{3},x_{4}) + 
\lambda \: g(x_{1},x_{2},x_{3},x_{4}) \Bigr\} = 0.
\end{equation}
This condition in conjunction with the hermiticity of $\hat{\rho}$ gives
rise to an eigenvalue equation,
\begin{equation}
\sum_{n=1}^{4} \mbox{Re}\bigl\{ M_{k,n} \bigr\} \: x_{n} = 
-\lambda \: x_{k}.
\end{equation}
The eigenvectors $(x_{1}^{j},x_{2}^{j},x_{3}^{j},x_{4}^{j})$ 
($j=1,2,3,4$) of this real, symmetric matrix are orthonormal since
$g=1$.  Inserting the eigenvectors into Eq.~(20) and making use of their
ortho-normalization results in $F =
\mbox{max}\{ \eta^{j} \}$, where $\eta^{j}=-\lambda^{j}$ are the
corresponding eigenvalues of this matrix.  It is convenient to renormalize
this expression so that it is $1$ for a maximally entangled state and $0$
for a separable state,
\begin{equation}
E(\hat{\rho}) = 2 \: \Bigl( \mbox{max} \{ \eta^{j} , 0 \} 
- \frac{1}{2} \Bigr),
\end{equation}
where $\eta^{j}$ are the eigenvalues of the matrix
$M_{n,m}=\mbox{Re}\{\langle \Phi^{n}|\hat{\rho}|\Phi^{m}\rangle\}$,
$|\Phi^{j}\rangle$ being the maximally entangled basis states defined in
Eq.~(1).  

\section{The relation between the fully entangled fraction and the
concurrence}

The fully entangled fraction, once measured, establishes
lower and upper bounds for the concurrence.  It has been proved that the
fully entangled fraction is a lower bound for the entanglement of
formation \cite{11,27} and therefore a lower bound for the concurrence
which is monotonically related to the entanglement of formation.  The
states which form the lower bound are given by a convex sum of a maximally
mixed state and an arbitrary pure state,
\begin{equation}
\hat{\rho}_{-} = \epsilon \frac{\hat{1}}{4} + (1-\epsilon)
|\psi_{pure}\rangle\langle\psi_{pure}|, 
\end{equation}
where $(0<\epsilon<1)$.  If the pure state is decomposed in a Schmidt
basis, $|\psi_{pure}\rangle = \bigl( \hat{U}_{A} \otimes \hat{U}_{B}
\bigr) \bigl( \cos(\theta/2)|00\rangle + \sin(\theta/2)|11\rangle \bigr)$ 
\cite{25}, the local unitaries will not contribute and we find that
$E(\hat{\rho}_{-}) = C(\hat{\rho}_{-}) =  
(1-\epsilon)\sin\theta-\epsilon/2$ (The maximum between this number and
zero is implicit).  

\begin{figure}[t]
\begin{center}
\epsfig{file=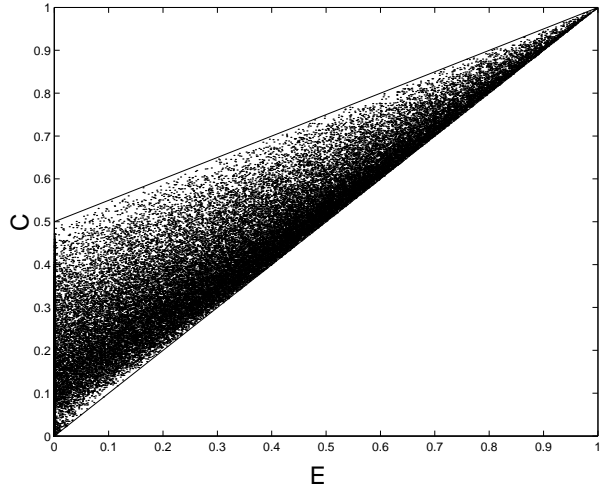,width=8cm}
\caption{Concurrence $C$ vs.\ $E$ for 100,000 random density matrices.  In
order to achieve a more uniform distribution and better demonstrate the
upper and lower bounds, we plot a modified distribution consisting of a
convex sum between Eq.~(25) and Eq.~(26) with a parameter that varies
between $0$ and $.5$.  The upper and lower bounds (solid lines) are given
by $E=2C-1$ and $E=C$
respectively.}
\end{center}
\end{figure}

The upper bound for the concurrence is found numerically by doing a
numerical search over one million random density matrices (See Fig.~2),
\begin{equation}
R = \frac{ T T^{\dag} }{ \mbox{Tr}\{ T T^{\dag} \} },
\end{equation}
where $T$ is a 4X4 matrix whose elements $T_{n,m} = t_{r} +
i t_{i}$ are determined by the random numbers $t_{r},t_{i}$ chosen
uniformly on the interval $\{0,1\}$.  We find that the upper
bound for the concurrence occurs for states that are a convex sum of a
direct product state $|uv\rangle$ and a maximally entangled state
$|\Phi\rangle$,
\begin{equation}
\hat{\rho}_{+} = \zeta |uv\rangle\langle vu| +
(1-\zeta) |\Phi\rangle\langle\Phi|,
\end{equation}
such that $\langle uv|\Phi\rangle = 0$ and $(0<\zeta<1)$.  Taking
$|uv\rangle = |01\rangle$ and $|\Phi\rangle = |\Phi^{1}\rangle$, we
compute $E(\hat{\rho}_{+})=1-2\zeta$ and $C(\hat{\rho}_{+}) = 1 - \zeta$,
which implies $E(\hat{\rho}_{+})=2C(\hat{\rho}_{+})-1$.  These bounds,
taken together, imply that a non-zero $E$ is a
necessary condition for nonzero concurrence, but not a sufficient
one.  These results are consistent with similar entanglement of
distillation results found by Bennett et. al. \cite{28}.  We see that this
operational measure determines the range of possible concurrence
values $C_{pos}$ for a mixed state,
\begin{equation}
E \leq C_{pos} \leq \frac{E+1}{2}.
\end{equation}

\section{Conclusions}

In conclusion, we have found that the fully entangled fraction can be used
as an inclusive measure of entanglement applications in the case of two
qubit states.  That is, $F>1/2$ guarantees that a mixed state can be used
to achieve, on average, ``classically impossible'' results in either dense
coding, teleportation, entanglement swapping, or quantum cryptography
(Ekert protocol); all two qubit quantum information processing
applications which have been experimentally demonstrated to date.  This
quantity has a simple closed-form expression for general two qubit states
given by the largest eigenvalue of the real part of the density matrix
expressed in a ``magic'' Bell basis.  Although it appears that the fully
entangled fraction is the ``best'' measure of entanglement applications in
the sense of being the most inclusive, we leave this question open.  It
could be conceived that there are other two qubit applications or
definitions of fidelity which have direct experimental consequences that
include the fully entangled fraction as a subset.  In which case it would
define a new inclusive measure of these entanglement applications which
sets the threshold for accomplishing classically inconceivable quantum
information tasks.  This quantity may be of more practical use than
entanglement for characterizing the quantum informations processing
ability of more complicated systems.  For example, dense coding
generalized to $d \times d$ systems allows Alice
to use a maximally entangled state $|\Phi^{1}\rangle$ to encode $d^{2}/2$
bits in $d^{2}$ orthogonol states $|\Phi^{i}\rangle =
(\hat{1}\otimes\hat{U}_{i})|\Phi^{1}\rangle$ by applying $d^{2}$ local
unitary operators $\hat{U}_{i}$ (where $i=1,...d^{2}$).  Replacing this
maximally entangled state with a general density operator and defining the
fidelity as in Sec.~2.1 as an average of the $d^{2}$ results gives,
\begin{eqnarray}
F_{DC} & = & \frac{1}{d^{2}} \sum_{i=1}^{d^{2}} \: 
\langle \Phi^{i}| (\hat{1}\otimes\hat{U}_{i}) \hat{\rho}
(\hat{1}\otimes\hat{U}_{i})^{\dagger} |\Phi^{i}\rangle \nonumber \\
& = & \langle \Phi^{1}|\hat{\rho}|\Phi^{1}\rangle.
\end{eqnarray}
Maximizing this over all local unitaries (this is the same as maximizing
over all maximally entangled states \cite{29}) we see that the maximum
fidelity of dense coding in this more general case is also given by the
fully entangled fraction,
\begin{equation}
F_{DC}^{max} = F.
\end{equation}
$F_{DC}^{max}=1$ when $|\Phi^{1}\rangle$ is maximally entangled and
$F_{DC}^{max}=1/d$ when $|\Phi^{1}\rangle$ is pure and
separable.  Horodecki et.\ al.\ \cite{19} also found a similar result for
the maximum teleportation fidelity,
\begin{equation}
F_{T}^{max} = \frac{Fd + 1}{d+1}.
\end{equation}
A general analytic expression for the fully entangled fraction for the
general mixed case in this system is not known, however, there are known
analytic results in the case of pure states \cite{30}.  It may also be
possible to generalize the association of the fully entangled fraction
with fidelities of quantum information tasks in multipartite systems.  

We would like to thank Tanmoy Bhattacharya and Bill Munro for useful
discussions.  DME would like to thank the Los Alamos Summer School for
support and this work was supported by the Los Alamos National Laboratory
LDRD program.

\end{document}